\documentclass[aps,prl,twocolumn,groupedaddress]{revtex4}
\usepackage{amssymb,amsfonts,amsmath,graphicx,color,mathptm,times}
\usepackage{mathrsfs,natbib}
\usepackage{amsmath,amssymb, amsfonts, bbm}
\usepackage{graphics,epsfig,color,subfigure}
\usepackage{hyperref}
\usepackage{verbatim}
\usepackage{eufrak}
\usepackage{natbib}

\newcommand{\be}{\begin{equation}}
\newcommand{\ee}{\end{equation}}

\def\bea{\begin{align}}
\def\ena{\end{align}}

\def\beqa{\begin{eqnarray}}
\def\enqa{\end{eqnarray}}

\newcommand{\half}{\textstyle{ \frac{1}{2}}} 
\hypersetup{
colorlinks=true, 
linkcolor=red,   
citecolor=blue,       
filecolor=magenta,    
urlcolor=blue         
}

\def\x{{\mathfrak x}}
\def\p{{\mathfrak p}}

\begin{document}

\title{Complexified path integrals, exact saddles and supersymmetry}
\author{Alireza Behtash$^{1,3}$}
\author{Gerald V. Dunne$^2$}
\author{Thomas Sch\"{a}fer$^1$}
\author{Tin Sulejmanpasic$^1$}
\author{Mithat \"{U}nsal$^{1,3}$}
\affiliation{
$^1$ Department of Physics, North Carolina State University, 
Raleigh, NC 27695, USA\\
$^2$ Department of Physics, University of Connecticut, 
Storrs, CT 06269, USA\\
$^3$ Department of Mathematics,  Harvard University, 
Cambridge, MA, 02138, USA}

\begin{abstract}
In the context of two illustrative examples from supersymmetric quantum 
mechanics we show that the semi-classical analysis of the path integral 
requires complexification of the configuration space and action, and the 
inclusion of complex saddle points, even when the parameters in the 
action are real.
We find new exact complex saddles, and show that without their contribution 
the semi-classical expansion is in conflict with basic properties such as 
positive-semidefiniteness of the spectrum, and constraints of supersymmetry.
Generic saddles are not only complex, but also possibly multi-valued, and  
even singular. 
This is in contrast to instanton solutions, which are real, smooth, and 
single-valued.  
The multi-valuedness of the action can be interpreted as a  hidden 
topological angle, quantized in units of $\pi$ in supersymmetric theories. 
The general ideas also  apply  to non-supersymmetric theories. 
\end{abstract}

\maketitle

{\noindent\bf Introduction:} 
We address the question of how to properly define the 
semi-classical expansion of the path integral in quantum mechanics and
quantum field theory. This question goes beyond
the problem of studying the semi-classical approximation, because the theory of 
resurgence shows that the semi-classical expansion encodes perturbative
as well as non-perturbative effects, and may provide a complete definition
of the path integral \cite{Argyres:2012ka,Dunne:2012ae}. We consider
a set of examples for which we show that the path integral measure
and action must be complexified, and that novel complex saddle
points appear.  The usefulness of complexification is not surprising from the
point of view of the steepest descent method for ordinary integration,  
but  important new effects  appear in functional integrals.  We show that in generic cases complexification is indeed essential. 
Our results go beyond proposals in the literature to 
complexify the path integral in cases where coupling constants are 
analytically continued away from their physical values, as described  in the work of Witten on Chern-Simons theory \cite{Witten:2010cx}, and Harlow, Maltz and Witten on Liouville 
theory \cite{Harlow:2011ny}, and is potentially related to the complexification of the 
phase space formulation of path integral \cite{Witten:2010zr}. Complex saddles 
were previously studied as a computational tool in quantum mechanics, see e.g. 
\cite{Balian:1974ah,Brezin:1976wa,Lapedes:1981tz,Balitsky:1985in}. Complex 
path integrals were also studied in connection with the sign problem in 
the Euclidean path integral of QCD and related model systems at finite chemical 
potential \cite{Cristoforetti:2012su,Fujii:2013sra,Aarts:2014nxa,Tanizaki:2015rda,Fujii:2015bua}.
 Here, we demonstrate the {\it necessity} of complexification even for the physical theory with real couplings. In \cite{BDSSU} we show that these complex saddles have a natural interpretation in terms of thimbles in Picard-Lefschetz theory.

There are several calculations in field theory that suggest the 
importance of complex saddle points. As an example consider  
${\cal N}=1$ supersymmetric gluodynamics on ${\mathbb R}^3\times S_1$ with 
SUSY preserving boundary conditions. 
This theory is confining, and has a non-perturbatively generated bosonic potential for 
the Polyakov line.  The 
potential for the Polyakov line can be computed using bions, 
molecules of monopole-instantons \cite{Unsal:2007jx,Poppitz:2012sw}. 
Bions also determine the vacuum energy, with the conclusion that 
supersymmetry is unbroken, e.g, for $SU(2)$ theory, 
$E_{\rm gr} \propto   - e^{ -2S_m} - e^{ -2S_m \pm  i   \pi} =0$ where the first 
is from magnetic bion and the latter from neutral bion. 
This calculation agrees with a calculation 
based on supersymmetry and the monopole instanton induced superpotential
\cite{Davies:1999uw}.   A puzzle concerning this result is that 
 the sum over  different bion types   give zero vacuum energy, despite the 
fact that contribution of real saddles is  universally  negative-semidefinite \footnote{This is the case when  the topological theta angle is set to zero.}.

 The calculations in  \cite{Poppitz:2012sw,Poppitz:2011wy} are based on analytic 
continuation in the coupling constant. Ref.~\cite{Behtash:2015kna} reinterprets  the relative sign between the two 
different bion types as a hidden topological angle (HTA), 
a factor $\exp(i\pi)$ associated with the relative phase in the quasi-zero mode  Lefschetz thimble, 
which is nothing but a  direction in field space.  This result  suggests that the calculation 
can be done directly for real values of 
$g$, and that  bions arise as exact (non-BPS) saddle point 
solutions of the complexified path integral, and furthermore that the HTA is 
related to the imaginary part of the complexified action. 

 SUSY gluodynamics on ${\mathbb R}^3\times S_1$ is not an isolated case. 
Similar phenomena occur in ${\cal N}=1$ $SU(2)$ SUSY QCD \cite{Yung:1987zp},  in three-dimensional SUSY 
gauge theory \cite{Affleck:1982as}, and in  ${\cal N}=2$ SUSY QM \cite{Behtash:2015kva}.   
In this paper we make 
the basic idea precise in the context of SUSY quantum mechanics.

\vspace{0.2cm}
{\noindent \bf Formalism and holomorphic Newton's equation:}
Consider the Euclidean quantum mechanical path integral as a sum over real paths,
$ Z= \int Dx (t) \;\exp(-\frac{1}{\hbar}S_E)$, with $S_E = \int dt ( \half 
\dot x^2 +V(x))$. The critical points   solve Newton's equation in the inverted potential, $\frac{d^2 x}
{dt^2} = + \frac{\partial V}{\partial x}$. This leads to the standard
multi-instanton calculus in quantum mechanics. More general saddle points 
appear in the  complexified path integral   
\be
Z =  \int_\Gamma  Dz \;   e^{ -\frac{1}{\hbar} {S}[z(t)]} \, , 
   \hspace{0.3cm}
   S[z(t)]= \int dt \left( \half \dot z^2  +   V(z)  \right) \, ,
\label{comp_pi}
\ee
where $\Gamma$ is an integration cycle that has the same dimensionality
as the original real path integral. The critical points of the complexified 
path integral  solve the {\it holomorphic 
Newton's equation} in the inverted potential $-V(z)$:
 $\frac{\delta{S}}{\delta z}=0 \Rightarrow   
   \frac{d^2z}{dt^2}  = + \frac{\partial V}{\partial z}\, .   $
In terms of real and imaginary parts of the potential,  $V(z) = V_{\rm r}
(x,y) +i V_{\rm i}(x,y)$, we get 
\begin{align}
\frac{d^2 x}{dt^2}  = + \frac{\partial { V_{\rm r}} }{\partial x}   \quad, \quad
\frac{d^2 y}{dt^2}  = - \frac{\partial { V_{\rm r}} }{\partial y}  ,
\label{h-Newton-ri}
\end{align}
where we have used the Cauchy-Riemann equations $\partial_x V_{\rm r} = 
\partial_y V_{\rm i}$, and $\partial_y V_{\rm r} = -\partial_x V_{\rm i}$.
An important aspect of \eqref{h-Newton-ri} is that it does {\bf not} describe 
an ordinary two-dimensional classical mechanical system:  the holomorphic  
classical mechanics  is not the same as the motion of 
a particle in the two-dimensional  inverted potential  $-V_{\rm r}(x,y)$. Instead
of the usual Newton equations with force $\vec{\nabla}V_{\rm r}(x,y)$, the 
force in the $x$-direction is due to $\nabla_x V_{\rm r}(x,y)$ while 
the force in the $y$-direction is due to $-\nabla_y
V_{\rm r}(x,y)$.  This has  
interesting consequences.

\begin{figure}[t] 
\includegraphics[width=7.5cm]{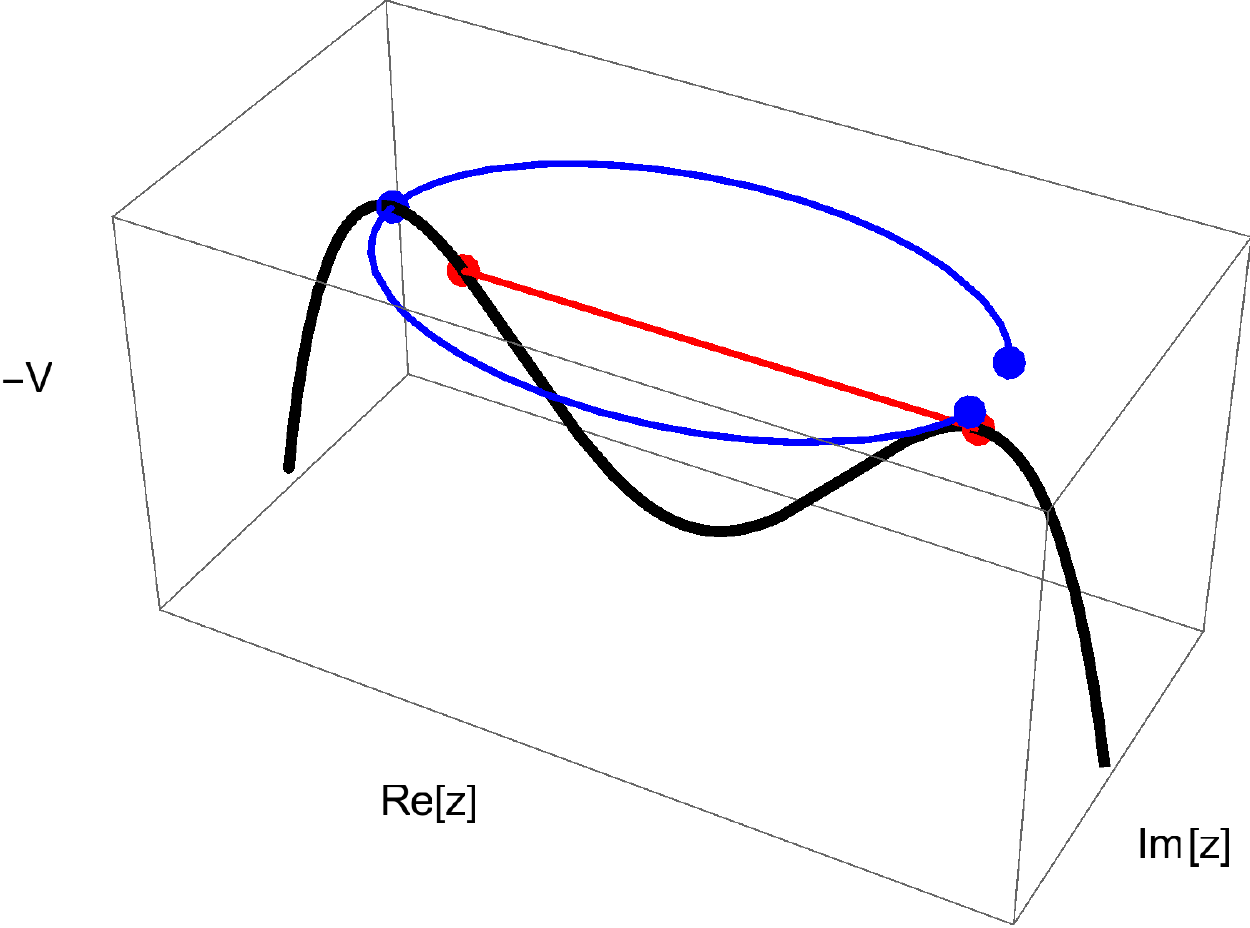}
\caption{Real and complex solutions in the inverted tilted double well
potential. The inverted potential (on the real axis) is shown in black, 
the real bounce and associated critical and turning points are shown in red, and the 
pair of complex bions and turning and critical  points are blue. The blue points 
correspond to $z_1^{\rm cr}$ and $z_T, z_T^*$ in \eqref{dw_cb}. Note that 
the motion takes place in the real and imaginary parts of the complex 
potential, as explained in the text. }
\label{fig_dw_tr}
\end{figure}

\vspace{0.2cm}
{\noindent\bf  Supersymmetric quantum mechanics:}
Consider supersymmetric quantum mechanics with the superpotential ${\cal W}(\x)$ 
\begin{equation}
S = \int dt \left( \half \dot \x^2  +  \half ({\cal W}')^2 
   +  \left[\bar \psi \dot \psi +   p {\cal W}''  \bar \psi \psi  \right]
   \right) \, ,
\label{SUSY-QM}
\end{equation}
corresponding to  $p=1$.  The parameter $p$ will  be used to deform the theory away from the supersymmetric point \cite{Balitsky:1985in}. 
We choose ${\cal W}(\x)$ with more than one critical point, so that there 
will be real instantons. By projecting to fermion number eigenstates one 
obtains a pair of Hamiltonians $H_{\pm}$ \cite{Witten:1982df}:
\begin{align} 
H_{\pm} =  \half \hat \p^2 +  V_{\pm} (\x) \, , \hspace{0.4cm} 
V_{\pm} (\x)=  \half ({\cal W}'(\x))^2   \pm   \tfrac{p}{2} {\cal W}''(\x)  \, . 
\label{SUSY-QM-B}  
\end{align}
In the following we consider superpotentials of the 
form ${\cal W}(\x)=\frac{1}{g}W(\sqrt{g}\x)$, and rescale $x=\sqrt{g}\x$. 
Then the Euclidean action takes the form $S_E=\frac{1}{g}\int dt(\half \dot{x}^2+V_{\pm}(x))$. 
We work with the bosonized
description \eqref{SUSY-QM-B}. Note that compared to the original 
bosonic potential $\half (W')^2$ the bosonized theory contains an 
$O(g)$ term that arises from integrating out the fermions. The quantum 
modified holomorphic equations of motion in the inverted potential   
$-V_{+}(z)$ is 
\begin{align} 
\frac{d^2 z}{dt^2} = W'(z)W''(z) 
+ \frac{p g}{2} W'''(z)  \, . 
\label{H-EOM-BOS}
\end{align}

\vspace{0.2cm}
{\noindent\bf  Double well potential:}
Consider  $W(x)= x^3/3-x$, so that $V(x)$ is an asymmetric double 
well potential with an $O(g)$ ``tilt''. The ground state energy of the system  
is zero to all orders in perturbation theory, but non-perturbatively
supersymmetry is spontaneously broken and the ground state energy is 
non-zero and positive \cite{Witten:1982df}. Note that the positivity of the ground state
energy is a consequence of the SUSY algebra, $H=\half\{Q,\bar{Q}\}$, where 
$Q$ and $\bar{Q}$ are the SUSY generators. 

 In the original formulation (\ref{SUSY-QM}) this can be understood as the 
contribution from {\it approximate} instanton-anti-instanton solutions of 
the bosonic potential $\half (W')^2$  \cite{Balitsky:1985in}. In the bosonized version we 
seek classical solutions in the inverted  potential $-V_+$. However, the real equations 
of motion in the inverted potential have no finite action configurations 
except for the trivial perturbative saddle, and an exact (real) bounce solution. But this bounce is related to the false vacuum and is not directly relevant for ground state properties, which are determined by saddles starting at the {\it global} maximum of the inverted potential. But the real motion of a classical 
particle starting at such a  global maximum is
unbounded, and  has infinite action. 

\begin{figure}[t] 
\includegraphics[width=8.5cm]{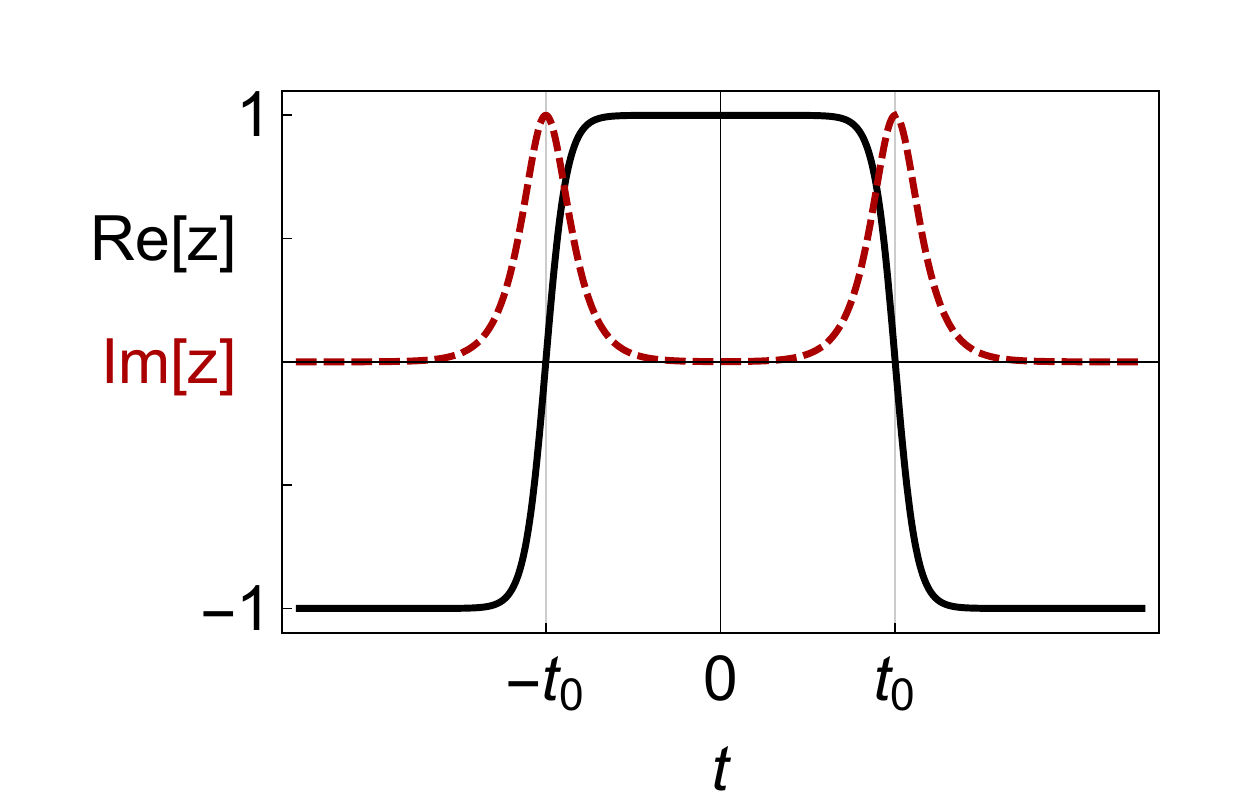}
\caption{Complex bion solution in supersymmetric quantum mechanics
with a double well potential. The black and red lines show the real
and imaginary part of the solution for $pg=1\cdot 10^{-6}$. The 
characteristic size of the solution is ${\rm Re}[2t_{0}]\simeq \half \log\frac{16}{pg}$. 
For larger values of $pg$ the two tunneling event merge. }
\label{fig_dw_cb}
\end{figure}

On the other hand, the holomorphic Newton's equation (\ref{H-EOM-BOS}) does support 
finite action solutions starting from the global maximum. There are {\it exact} 
finite action complex solutions that  start at the global maximum of the inverted 
potential and bounce back from one of the two complex turning points, whose real 
part is located  near 
   the top of the local maximum, see Fig.~\ref{fig_dw_tr}. 
 We refer to this as the ``complex bion'' solution:
\begin{eqnarray}
\label{dw_cb}
z_{\rm cb}(t) &=&z_1^{\rm cr}-\frac{z_1^{\rm cr}-z_T}{2} \coth\left(\frac{\omega_{\rm cb} t_0}{2}\right)\left[\tanh\left(\frac{\omega_{\rm cb}(t+t_0)}{2}\right)\right.\nonumber\\
&&\left. - \tanh\left(\frac{\omega_{\rm cb}(t-t_0)}{2}\right)\right]\, ,
\end{eqnarray}
where $z_{\rm cb}(\pm\infty)=z_1^{\rm cr}$ is the global maximum of the 
inverted potential, and $z_T=-z_1^{\rm cr}\pm i\sqrt{pg/(-z_1^{\rm cr})}$ are the 
complex turning points.  
$\omega_{\rm cb}=\sqrt{V''(z_1^{\rm cr})}$ is the natural frequency at $z_1^{\rm cr}$, 
and  the complex parameter  $t_0$ is 
\be 
 t_0 =\frac{2}{\omega_{\rm cb}} {\rm arccosh} 
 \left[
  \frac{3\omega_{\rm cb}^2}
       {\omega_{\rm cb}^2-V''(z_1^{\rm cr})}\right]^{1/2} 
   \approx \frac{1}{2\omega_{\rm cb}}\ln \left(-\frac{16}{p g} \right)
\ee
where ${\rm Re}[2t_0]$ is the complex bion size. 

It is straightforward to verify that \eqref{dw_cb}
is a solution to the holomorphic equation of motion. The solution 
is shown in Fig.~\ref{fig_dw_cb}. The real part
resembles an instanton-anti-instanton pair with size 
$\ln\frac{16}{pg}$,   and  the action is
\begin{eqnarray}
S_{\rm cb} \simeq \left( \frac{8}{3g} +p\, \ln \frac{16}{p g}+\dots\right)  
  \pm i\, p\, \pi\, , 
\end{eqnarray}
whose real part  is slightly larger than two-instanton  action, $2S_I$. 
The sign of the imaginary part ${\rm Im} S_{cb}=\pm p\pi$ corresponds to 
the choice between two complex conjugate saddles.

The imaginary part of the action is defined modulo $2\pi$, so the choice between the two complex conjugate saddles does not lead to an 
ambiguity in the amplitude for $p = 1$.
However, the factor $e^{i\pi}$ 
is related to a hidden topological angle which is crucial to obtain 
the correct sign for the ground state energy. In the semi-classical
limit the ground state can be understood as a  dilute gas of 
complex bions: 
\be 
 E_{\it gs} \sim -e^{\pm i\pi} e^{-2S_I} \sim + e^{-2S_I} >0 \, , 
\ee
in agreement with  known results \cite{Salomonson:1981ug, Balitsky:1985in}. 
The bosonized description makes the most  crucial point  clear.  From a semi-classical view point, the positivity of the ground state energy 
and hence,  consistency with the supersymmetry algebra owes  its existence 
to the complexity of the exact solution,  and  to the    hidden topological angle associated with it.

This complex bion solution can also be constructed by analytic continuation of the
real bounce solution. To this end, we consider analytic 
continuation in the parameter $p\rightarrow pe^{i \theta}$, with corresponding 
potential
\begin{align} 
V_{\theta} (x)= \half (W'(x))^2  + \frac{pe^{i\theta}g}{2}  W''(x) \, .  
\label{V_theta}  
\end{align}
Only  $\theta=0,\pi$ are physical theories,
but the continuous $\theta$ parameter is useful in order to understand the 
relation between different saddle points. 
The regular bounce solution starts at the local (smaller) maximum of the 
inverted potential, and gets reflected at a turning point below the global (larger) maximum. 
This solution is described by an ordinary elliptic integral. The analytic continuation, $p\to pe^{i\theta}$, produces a complex solution with finite action solution at any 
$\theta$, and can be continued all the way to $\theta=\pi$, where the
local and the global maxima of the inverted potential are interchanged. 
The solution comes back to itself after $4\pi$ rotation, it has order 2  monodromy. 
 At $\theta=\pm \pi$ we obtain exactly the complex conjugate pair of ``complex bion'' solutions (\ref{dw_cb}).

\vspace{0.2cm}
{\noindent\bf  Quantization of  hidden topological angle in supersymmetric theory:}
In supersymmetric theories, since ground state energy is zero to all orders in perturbation theory, 
to avoid an   ambiguity in the ground state energy, it is essential that the imaginary 
part of the complex action is a multiple of $\pi$. Here, we will give a simple proof of this fact 
for the double well potential, which extends easily to the periodic potential. 
There are two complex bion solutions with 
complex conjugate turning points, and complex conjugate actions. 
This means that the imaginary part of the action can be computed from
the difference of the action of the two complex bions, ${\rm Im}\,S_{\rm cb}
=\pm \frac{1}{2i} (S_{\rm cb}^1-S_{\rm  cb}^2)$. We also note that the action is 
computed as a line integral of the quantity $\sqrt{2(E+V)}$ along the 
branch cut that connects the turning points of the solution. Here, 
$E$ is the energy of the solution in the inverted potential. This 
implies that the imaginary part of the action can be written as 
\be 
{\rm Im}\,S_{\rm cb} = \frac{1}{2g}\oint_C dz 
  \sqrt{2E+(W')^2+pg W''}\, ,
\ee
where the contour $C$, which arises from first going around the 
branch cut connecting $z_1^{\rm cr}$ and $z_T$, and then around
the cut connecting $z_1^{\rm cr}$ and $z_T^*$, encircles all the 
points $z_1^{\rm cr}, z_T,z_T^*$. This implies that we can deform the contour 
into a large circle in the complex $z$-plane. If $W$ grows as 
a positive power of $z$ we have $(W')^2\gg W''$, so that the 
integrand can be expanded in powers of $E/(W')^2$ and $W''/(W')^2$. 
The first term is a total derivative, and the second term vanishes 
because its residue is zero. Terms of second order 
and higher in $1/(W')$ vanish faster than $1/z$ as $z\to \infty$. 
The only contribution comes from
\be 
{\rm Im}\,S_{\rm cb} = \frac{p}{4}\oint dz \frac{W''}{W'}
    = \frac{p}{2}\oint \frac{dW'}{W'} =i p\pi\, ,
\ee
where we used the fact $W'\sim z^2$ winds twice as $z$ 
encircles the critical points.  This proves the quantization of the HTA in the supersymmetric $p=1$  limit. 

For the $p\neq 1$ non-supersymmetric deformation of the theory, the perturbative  ground state energy does not vanish anymore, but the energy spectrum must still be unambiguous.  
In that case, we show in \cite{BDSSU} that  the ambiguity 
inherent to the Borel resummation of perturbation theory cancels exactly the  two-fold ambiguous complex bion amplitude, as an explicit illustration of resurgence. 


\vspace{0.2cm}
{\noindent\bf  Periodic potential:}
Now consider the superpotential $W(x)=4\cos(x/2)$. In
this system supersymmetry is unbroken \cite{Witten:1982df}. There are two degenerate
ground states, one bosonic and one fermionic, both with vanishing
ground state energy. After the fermion is integrated out we obtain
the bosonic potential
\be 
\label{v_sg}
 V_{\pm}(x) = 2\sin^2(x/2) \pm \frac{p g}{2}\cos(x/2)\, . 
\ee
\begin{figure}[t]
\centering
\includegraphics[width=7cm]{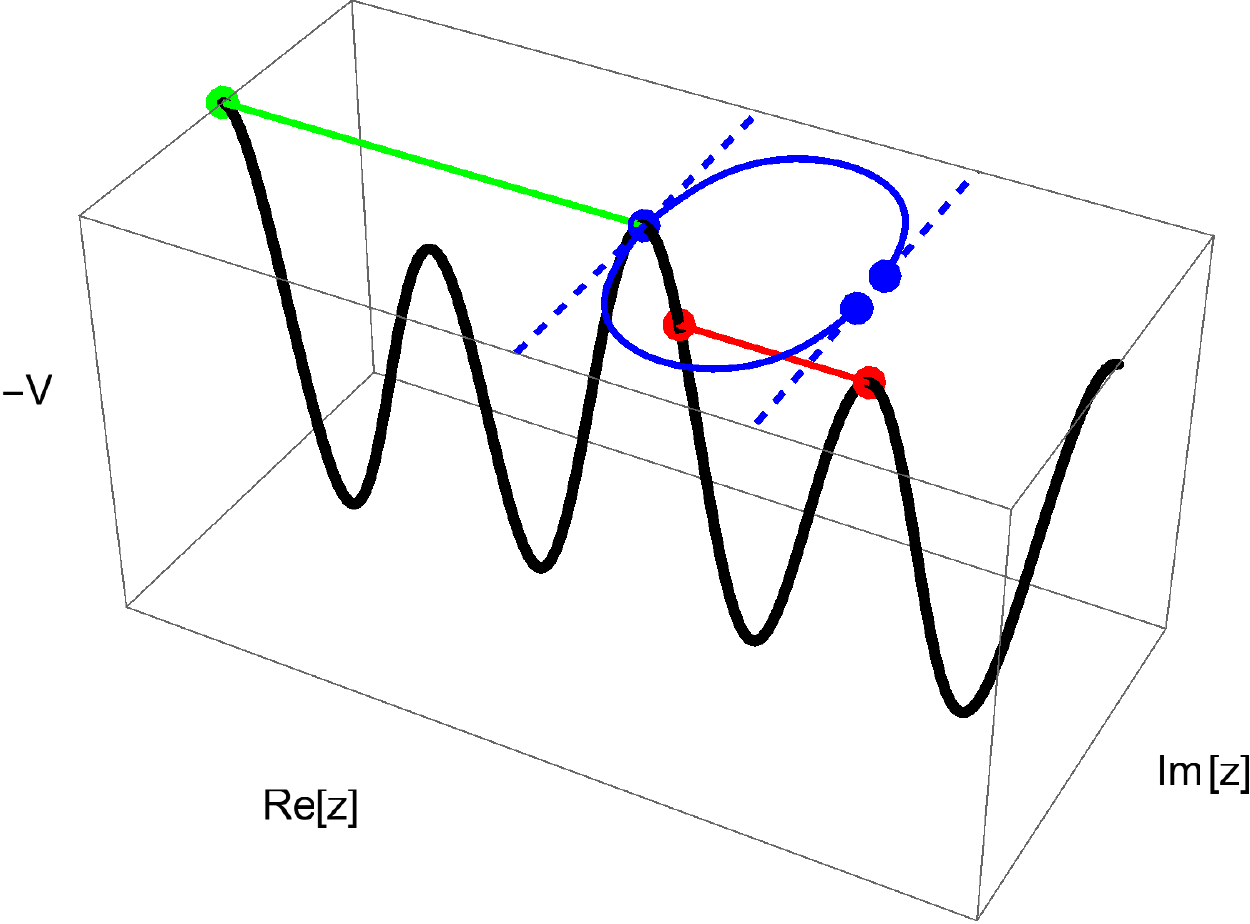}
\caption{
Real and complex solutions in the quantum modified inverted Sine-Gordon potential. 
The inverted potential (on the real axis) is shown in black, the real bounce and 
associated critical and turning points are shown in red, the pair of complex bions 
and associated turning as well as critical points are blue, and the real bion is 
shown in green. In order to smoothen the (singular) complex bion the solution is 
plotted at $\theta= 0.95 \pi$. The singular limit is shown as the dashed line.
Note that the vacuum properties are governed by the real and complex bion solutions.
}
\label{fig:periodic_cartoon}
\end{figure}
The inverted 
potential, Fig.~\ref{fig:periodic_cartoon},   has  global maxima at $x=4 n \pi$, and  local maxima at 
$x=(4n+2)\pi$. (The potential has period $4\pi$). There is an exact real 
bounce solution starting at the local maximum, and bouncing from a real turning point, 
but again this is not directly relevant for ground state properties. 
Now we find two types of exact bion solutions, shown in Fig.~\ref{fig:periodic_cartoon}. The first is a ``real bion'',
connecting neighboring global maxima, say at $x=0$ and $x=4\pi$. It has the form of an instanton-instanton solution, and as such has no analogue in the double-well case. 
There is also a {\it complex} bion solution that starts from a global maximum of 
the inverted potential, and is reflected from a complex turning point, 
with  real part near  the local maximum. This solution can be found directly 
or by analytic continuation from the real bounce, $p\to p e^{i\theta}$, and leads to an exact finite action complex saddle: 
\be
\label{z_cb_sg}
z_{\rm cb}(t) = 2\pi\pm 4 \, \left({\rm arctan} \, e^{-\omega_{\rm cb} (t-t_0)} +{\rm arctan} \, e^{\omega_{\rm cb} (t+t_0)} \right)\, , 
\ee
where $\omega_{\rm cb}= \sqrt  {V'' (0 ) }  =  \sqrt{1 +\tfrac{pg}{8}}$.  The complex parameter 
 $t_0\simeq \frac{1}{2 \omega_{\rm cb}} \ln \left(-\frac{32}{p g} \right)$, 
 where ${\rm Re}[2t_0]$ is the complex bion size.
 The action is
\begin{eqnarray}
S_{\rm cb} \simeq \left(\frac{16}{g} + p\, \ln \frac{32}{p g} +\dots\right) 
 \pm i\, p\, \pi \, . 
\end{eqnarray}
The complex bion has the form of a complex instanton/anti-instanton molecule.
An interesting new feature of this solution is that it is singular at $t=\pm t_0$, even though the action is finite. Physically this is because the real part of the holomorphic potential has ridges along the $y$ direction, and the holomorphic equations of motion allow the particle to {\it roll up}  (notice  relative signs in (\ref{h-Newton-ri}))  along one ridge and then jump to the next ridge at infinity before rolling back again. 

\begin{figure}[t] 
\includegraphics[width=8cm]{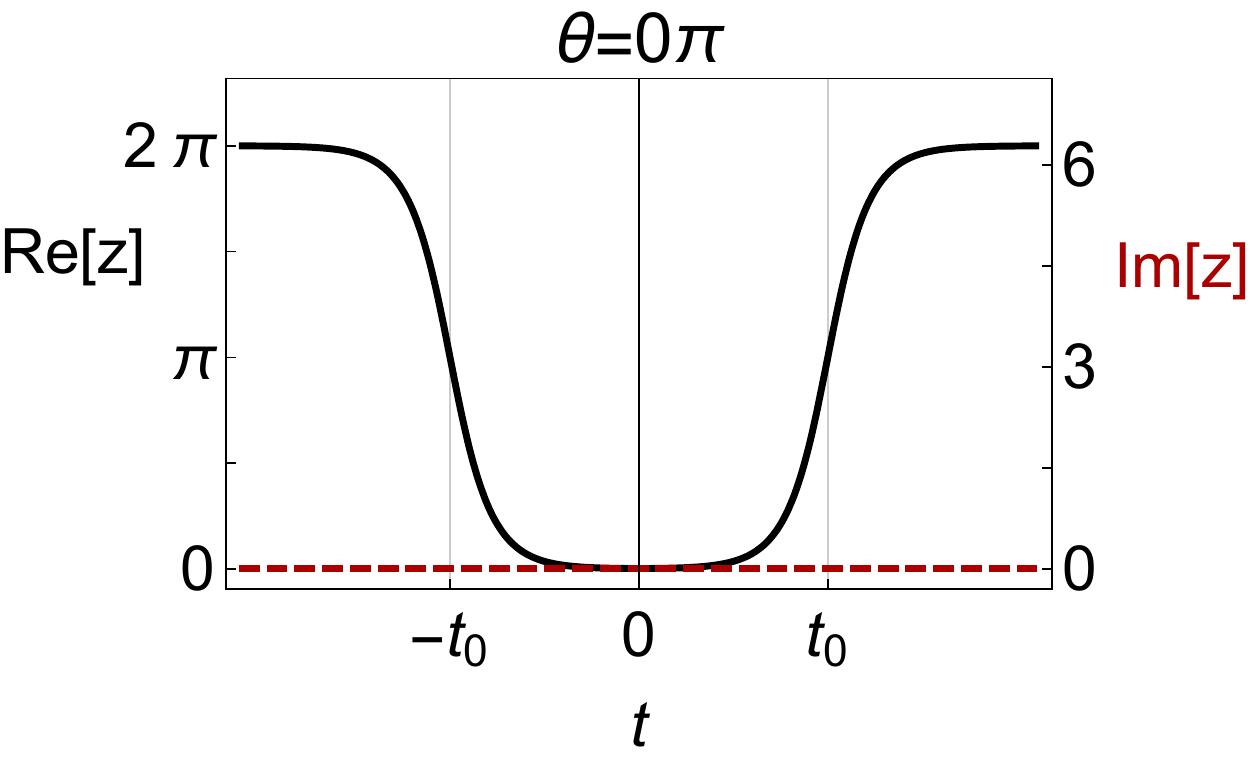}
\includegraphics[width=8cm]{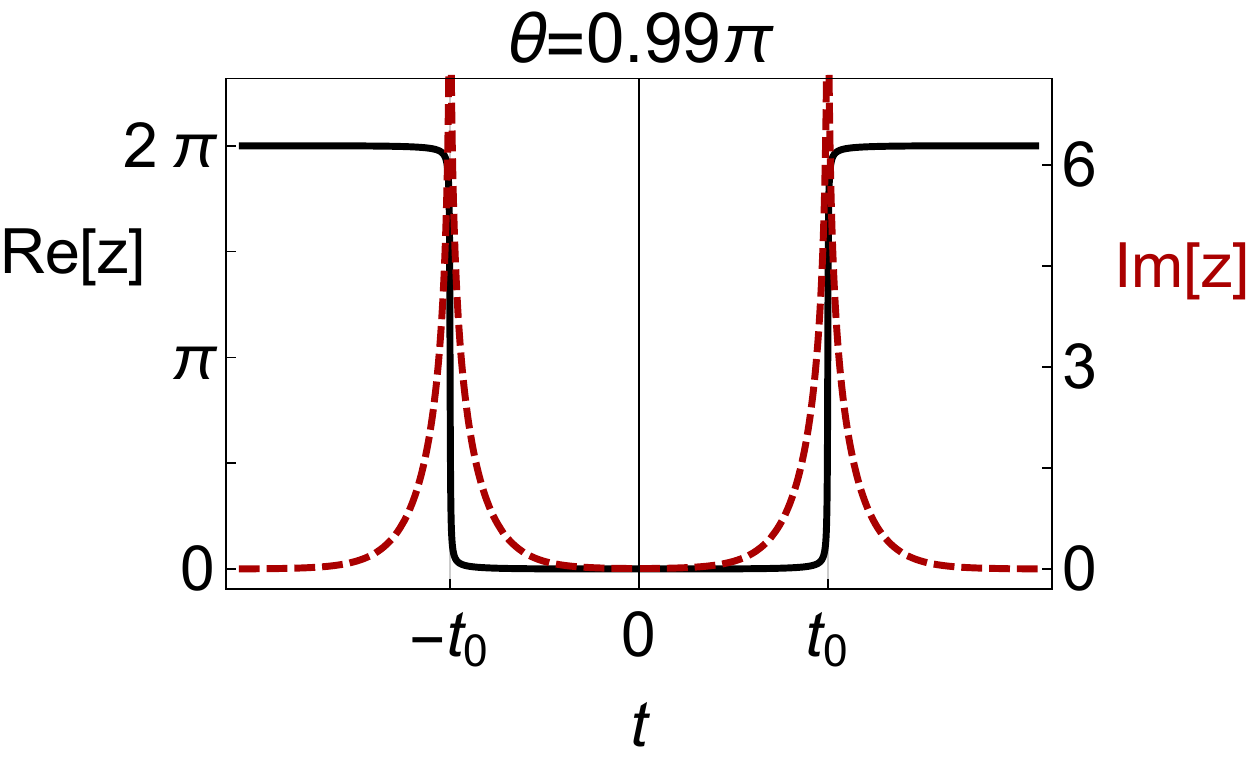}
\caption{Complex solutions in the quantum modified Sine-Gordon potential
with $p g=2\cdot 10^{-5}$. $\theta=0$ correspond to the real bounce solution. 
At $\theta=\pi^{-}$, the real bounce turns into a complex bion. The 
characteristic size of the solution is ${\rm Re}[2t_{0}] \approx \ln\frac{32}{pg} $. }
\label{fig_sg_cb}
\end{figure}

\begin{figure}[t]
\centering
\includegraphics[width=8cm]{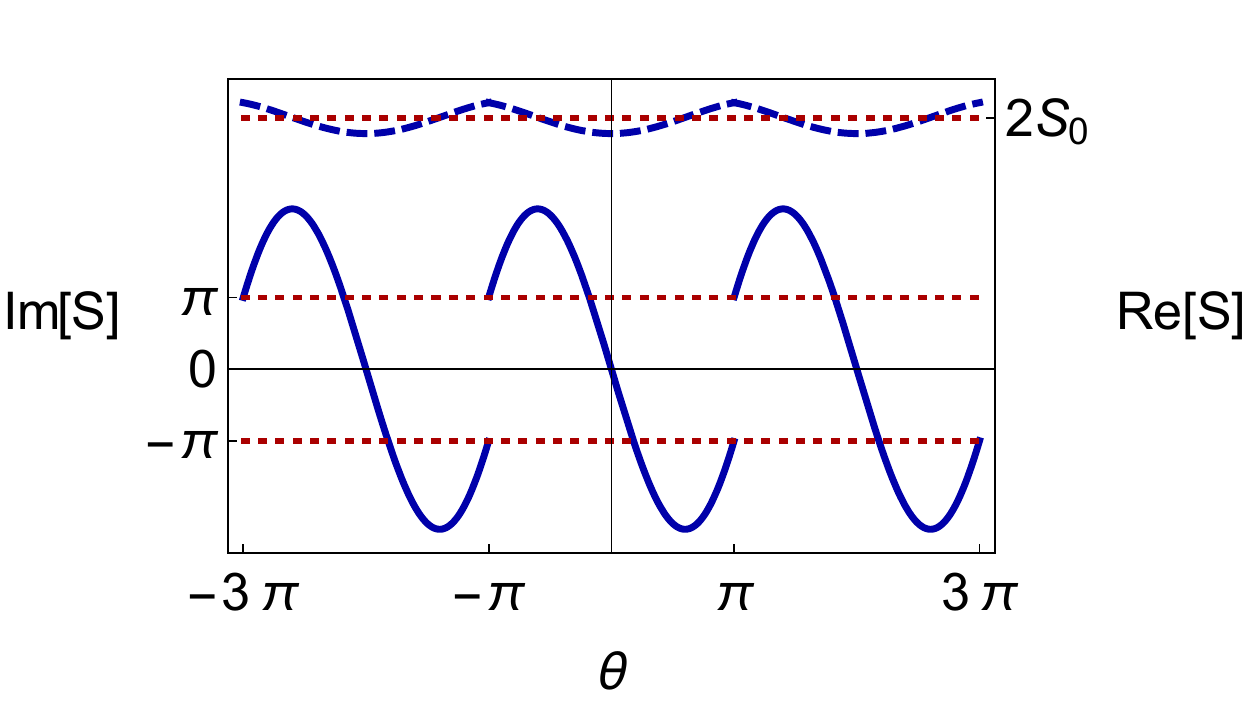}
\caption{Action of the complex saddle solution in the Sine-Gordon 
potential as a function of $\theta$ for $p g=0.1$. $\theta=0$ corresponds 
to the real bounce, and $\theta=\pi$ is the complex bion, a multi-valued 
singular solution. }
\label{fig_sg_act}
\end{figure}

 The analytic continuation in $\theta$ smooths this singularity, and the solution is correspondingly multivalued as $\theta\to \pi
\pm \epsilon$: see Fig.~\ref{fig_sg_cb}. As $\theta\to \pi$ the real part 
of $z_{\rm cb}(t)$ has a discontinuity, and the imaginary part
diverges. 
The action is finite, because the 
divergence in the action integral due to the singular behavior
in ${\rm Re}\,z(t)$ and ${\rm Im}\,z(t)$ cancel. Fig.~\ref{fig_sg_act}  shows the real and imaginary parts of the action as a function of the $\theta$ parameter. 

 In the semi-classical limit the ground state can be described 
as a dilute gas of complex and real bions, with 
energy
\be 
 E_{\it gs} \sim -e^{-S_{\rm cb}} - e^{-S_{\rm rb}}
  = -e^{\pm i\pi}e^{-2S_{\rm rb}}-e^{-2S_{\rm rb}} = 0 \, , 
\ee 
consistent with the requirement of supersymmetry.  
The non-inclusion of the 
multi-valued saddle would result in  a negative ground state energy and a conflict with  the constraints of supersymmetry algebra.
This proves that 
in order for the   semi-classical analysis to be consistent with  the supersymmetry algebra, it is essential to include
 singular, multi-valued  complex bion solution. This  resolves 
a deep puzzle raised in  \cite{Harlow:2011ny}.

\vspace{0.2cm}
{\noindent\bf  Conclusions:}
We have presented two examples that demonstrate the
need to include complex, and even singular and multi-valued, saddle 
point solutions of the path integral. We obtained exact finite action 
saddle points of the complexified path integral in supersymmetric 
quantum mechanics with a double well and Sine-Gordon potential. In 
both cases these new complex bion configurations are essential in order to obtain 
agreement with known results and the requirements of supersymmetry. 
This phenomenon is not restricted to quantum mechanics: 
analogous effects occur in several field 
theories, such as  sigma models with fermions \cite{Dunne:2012ae, Dunne:2012zk, 
Cherman:2014ofa, Misumi:2014bsa,Misumi:2014jua}
SUSY gluodynamics and QCD(adj) on $R^3\times S_1$, ${\cal N}=1$ \cite{Unsal:2007jx,  Anber:2011de,
Misumi:2014raa}, 
$SU(2)$ SUSY QCD with one  
quark flavor \cite{Yung:1987zp,Affleck:1983rr}, and three dimensional SUSY  ${\cal N}=2$ gauge 
theory \cite{Affleck:1982as}.   Clearly, it is of interest to study these field theories, 
and ultimately QCD, using complexified path integrals.

\vspace{0.2cm}
{\noindent\bf Acknowledgments:}
We thank P. Argyres, D. Harlow, and  E. Witten for useful comments and discussions. 
M.~{\"U}.was partially supported by the Center for Mathematical Sciences and 
Applications (CMSA) at Harvard University.
We acknowledge support from DOE grants DE-FG02-03ER41260 and DE-SC0010339.
 
\bibliographystyle{apsrev4-1}
\bibliography{bibliography}

\end{document}